\documentstyle[twoside,fleqn,epsfig]{article}

\def\be{\begin{equation}}
\def\ee{\end{equation}}
\def\bea{\begin{eqnarray}}
\def\eea{\end{eqnarray}}
\newcommand{\lsim}{\mathrel{\mathop{\kern 0pt \rlap
  {\raise.2ex\hbox{$<$}}} \lower.9ex\hbox{\kern-.190em $\sim$}}}
\newcommand{\gsim}{\mathrel{\mathop{\kern 0pt \rlap
  {\raise.2ex\hbox{$>$}}}
  \lower.9ex\hbox{\kern-.190em $\sim$}}}

\newcommand{\AmS}{{\protect\the\textfont2
  A\kern-.1667em\lower.5ex\hbox{M}\kern-.125emS}}
\normalsize
\begin{document}

\baselineskip=0.65cm

\begin{center}
\Large
{\bf Model independent result on possible diurnal effect in DAMA/LIBRA--phase1}
\rm
\end{center}

\large

\begin{center}

R.\,Bernabei$^{a,b}$,~P.\,Belli$^{b}$,~F.\,Cappella$^{c,d}$,~V.\,Caracciolo$^{e}$,~S.\,Castellano$^{e}$, 
\vspace{1mm}

R.\,Cerulli$^{e}$,~C.J.\,Dai$^{f}$,~A.\,d'Angelo$^{c,d}$,~S.\,d'Angelo$^{a,b}$,
\vspace{1mm}

~A. Di Marco$^{a,b}$,~H.L.\,He$^{f}$, A.\,Incicchitti$^{d}$,~H.H.\,Kuang$^{f}$,
\vspace{1mm}

~X.H.\,Ma$^{f}$,~F.\,Montecchia$^{b,g}$,~D.\,Prosperi$^{c,d,}\footnote{Deceased}$,
\vspace{1mm}

~X.D.\,Sheng$^{f}$,~R.G.\,Wang$^{f}$,~Z.P.\,Ye$^{f,h}$
\vspace{1mm}

\normalsize
\vspace{0.4cm}

$^{a}${\it Dip. di Fisica, Universit\`a di Roma ``Tor Vergata'', I-00133  Rome, Italy}
\vspace{1mm}

$^{b}${\it INFN, sez. Roma ``Tor Vergata'', I-00133 Rome, Italy}
\vspace{1mm}

$^{c}${\it Dip. di Fisica, Universit\`a di Roma ``La Sapienza'', I-00185 Rome, Italy}
\vspace{1mm}

$^{d}${\it INFN, sez. Roma, I-00185 Rome, Italy}
\vspace{1mm}

$^{e}${\it Laboratori Nazionali del Gran Sasso, I.N.F.N., Assergi, Italy}
\vspace{1mm}

$^{f}${\it Institute of High Energy Physics, Chinese Academy of Sciences, P.O. Box 918/3, Beijing 100049, China}
\vspace{1mm}

$^{g}${\it Dip. di Ingegneria Civile e Ingegneria Informatica, Universit\`a di Roma ``Tor Vergata'', I-00133  Rome, Italy}
\vspace{1mm}

$^{h}${\it University of Jing Gangshan, Jiangxi, China}

\end{center}
	
\normalsize

\begin{abstract}
The results obtained in the search for possible diurnal effect in the {\it single-hit} 
low energy data collected by DAMA/LIBRA--phase1 (total exposure: 1.04 ton $\times$ yr)
deep underground at the Gran Sasso National Laboratory (LNGS) of the I.N.F.N. are presented.
At the present level of sensitivity the presence of any significant diurnal variation and of 
diurnal time structures in the data can be excluded for both the cases of solar and sidereal time. 
In particular, the diurnal modulation amplitude expected, because of the Earth diurnal motion,
on the basis of the DAMA Dark Matter annual modulation results is below the present sensitivity.
\end{abstract}

\vspace{5.0mm}

{\it Keywords:} Scintillation detectors, elementary particle processes, Dark 
Matter

\vspace{2.0mm}

{\it PACS numbers:} 29.40.Mc - Scintillation detectors;
                    95.30.Cq - Elementary particle processes;
                    95.35.+d - Dark matter (stellar, interstellar, galactic, and cosmological).

\section{Introduction}

The present DAMA/LIBRA \cite{perflibra,modlibra,modlibra2,modlibra3,pmts,mu,review,alllist} 
experiment, as the former DAMA/NaI \cite{alllist,RNC,ijmd,allDM4},
has the main aim to investigate the presence of Dark Matter (DM) particles in the galactic halo by exploiting
the model independent DM annual modulation signature (originally suggested in Ref. \cite{Drukier,Freese}). 
In particular, they have cumulatively reached a model independent evidence at 9.3 $\sigma$ C.L. for the presence
of DM particles in the galactic halo on the basis of the exploited DM annual modulation signature \cite{modlibra3}.

In the present work the DAMA/LIBRA--phase1 data (total exposure: 1.04 ton $\times$ yr)
are analysed in terms of possible diurnal variation of the rate of the 
{\it single-hit} events\footnote{i.e. those events where only one of the 25 detectors in DAMA/LIBRA fires; that is, each detector 
has all the others as anti-coincidence.} in low energy regions, both where the DM annual modulation signal is observed ($2-6$ 
keV, see Ref. \cite{modlibra,modlibra2,modlibra3} and references therein) and in the region just above ($6-14$ keV) for comparison.
Actually a diurnal effect with the sidereal time is expected for DM because of Earth rotation.
In Sect. 2 the diurnal modulation of the DM signal as a function of the sidereal time 
due to Earth rotation velocity contribution will be discussed and some relevant formulae will be presented;
this effect is model-independent and has several requirements as the DM annual modulation effect does. 
Thus, in Sect. 4 the data have been analyzed using 
the sidereal time referred to Greenwich, also sometimes called GMST.
Since potential environmental backgrounds can be in principle correlated with the solar time,
the analysis has been also performed in terms 
of solar time referred to the LNGS site.

\section{Expectation for DM diurnal effect because of the Earth rotation}
\label{s:diurnal}

Let us now introduce an interesting model independent effect which can 
induce a diurnal variation of the counting rate of the {\it single-hit} events: 
the diurnal modulation of the DM signal as a function of the sidereal time due to Earth rotation velocity contribution \cite{Diu09}.
As explained below, this effect is linked with the DM model independent annual modulation 
signature (see also for example Refs. \cite{dama_diu,lisa12,free13,mcca13}). 
To explain this point in this section we describe 
all the components of the motion of a
detector placed in a terrestrial laboratory with respect to an observer fixed in the Galactic frame, including both the
revolution of the Earth around the Sun and the rotation of the Earth 
around its axis.

As known, the velocity of the detector in the terrestrial laboratory can be expressed as following:
\begin{equation}
\vec{v}_{lab}(t) = \vec{v}_{LSR} + \vec{v}_{\odot} + \vec{v}_{rev}(t) + \vec{v}_{rot}(t), 
\label{eq:velearth}
\end{equation}
where $\vec{v}_{LSR}$ is the velocity of the Local Standard of Rest (LSR) due to the rotation of the Galaxy;
$\vec{v}_{\odot}$ is the Sun peculiar velocity with respect to LSR;
$\vec{v}_{rev}(t)$ is the velocity of the revolution of the Earth around the Sun and 
$\vec{v}_{rot}(t)$ is the velocity of the rotation of the Earth around its axis at the latitude and longitude of the laboratory.
Using the Galactic coordinate frame (that is $x$ axis towards the Galactic center, $y$ axis following the rotation of the Galaxy
and the $z$ axis towards the Galactic North pole), we have $\vec{v}_{LSR}=(0,v_0,0)$, with $v_0 = (220\pm50)$ km/s 
\cite{allDM4,astr_v0} (uncertainty at 
90\% C.L.), and $\vec{v}_{\odot} = (9, 12, 7)$ km/s \cite{delh65}, while the revolution and rotation velocities of the Earth depend on 
the sidereal time, $t$.

{\it Revolution of the Earth around the Sun.} This motion can be easily worked out
using the ecliptic coordinate system
$(\hat{e}^{ecl}_1, \hat{e}^{ecl}_2, \hat{e}^{ecl}_3)$,
where the $\hat{e}^{ecl}_1$ axis is directed towards the vernal equinox and $\hat{e}^{ecl}_1$
and $\hat{e}^{ecl}_2$ lie on the ecliptic plane. The right-handed convention is used.
In the Galactic coordinates, we can write\footnote{
The coordinates of these versors are firstly worked out in the equatorial coordinate system
by using the routines given in Ref. \cite{starlink} and then in the 
Galactic coordinate system by using the $RA$ and the $DE$ of the Galactic North pole
and of the Galactic center (see also later).}:
\begin{eqnarray}
\hat{e}^{ecl}_1 & = & (-0.05487,  0.49411, -0.86767), \nonumber \\
\hat{e}^{ecl}_2 & = & (-0.99382, -0.11100, -0.00035), \nonumber \\
\hat{e}^{ecl}_3 & = & (-0.09648,  0.86228,  0.49715). 
\end{eqnarray}
The ecliptic plane is tilted with respect to the galactic plane by $\approx 60^o$,
as $\hat{e}^{ecl}_3 \cdot (0,0,1) = 0.49715$.

The motion of the Earth in the ecliptic plane can be described as:
\begin{equation}
\vec{v}_{rev}(t) = V_{Earth}(\hat{e}^{ecl}_1\sin\lambda(t) - \hat{e}^{ecl}_2\cos\lambda(t))
\label{eq:velearth2}
\end{equation}
where $V_{Earth}$ is the orbital velocity of the Earth, which has a weak dependence on time 
due to the ellipticity of the Earth orbital motion around the Sun; its value ranges
between 29.3 km/s and 30.3 km/s.
For most purposes it can be assumed constant and equal to its mean value $\simeq$ 29.8 km/s.
On the other hand, when more accurate calculations are necessary, the routines in Ref. \cite{starlink} 
can be used:
they also take into account the ellipticity of the Earth orbit and the gravitational influence of
other celestial bodies (Moon, Jupiter, and etc.)\footnote{For
completeness, we remind the discussion about the ellipticity of the Earth orbit in Ref. \cite{mcca13},
which overcomes the description given in Ref. \cite{lsmi96}}.
Moreover, the phase in eq. \ref{eq:velearth2} can be written as $\lambda(t) = \omega (t-t_{equinox})$;
here $\omega = 2\pi/T$ with $T$ = 1 y, $t$ is the sidereal time and $t_{equinox}$ 
is the spring equinox time ($\approx$ March 21).

{\it Rotation of the Earth around its axis.} The simplest way to express this  motion is 
in the equatorial coordinate system $(\hat{e}^{ecs}_1, \hat{e}^{ecs}_2, \hat{e}^{ecs}_3)$,
where the $\hat{e}^{ecs}_1$ axis is directed towards the vernal equinox and $\hat{e}^{ecs}_1$
and $\hat{e}^{ecs}_2$ are on the equatorial plane; the $\hat{e}^{ecs}_3$ axis is towards the North pole. 
The right-handed convention is used.
To work out the Galactic coordinates of these versors, we use the equatorial coordinates of the
Galactic North pole: $RA = 192^o.859508$ and $DE = 27^o.128336$ ($RA$ is the Right Ascension and 
$DE$ is Declination); and of the Galactic center: $RA = 266^o.405100$ and $DE = -28^o.936175$,
evaluated at the Epoch J2000.0.
In the galactic coordinates, these versors can be written as:
\begin{eqnarray}
\hat{e}^{ecs}_1 & = & (-0.05487,  0.49411, -0.86767), \nonumber \\
\hat{e}^{ecs}_2 & = & (-0.87344, -0.44483, -0.19808), \nonumber \\
\hat{e}^{ecs}_3 & = & (-0.48384,  0.74698,  0.45599).
\end{eqnarray}
Therefore, we can write: 
\begin{equation}
\vec{v}_{rot}(t) = -V_{r}(\hat{e}^{ecs}_1\sin\delta(t) - \hat{e}^{ecs}_2\cos\delta(t))
\label{eq:velearth3}
\end{equation}
where $V_{r}$ is the rotational velocity of the Earth at the given latitude, $\phi_0$, of the laboratory:
$V_{r} = V_{eq} cos\phi_0$. The equatorial rotational velocity, $V_{eq}$, is equal to 0.4655 km/s.
Hence, at LNGS ($\phi_0 = 42^o27'$N and longitude $\lambda_0 = 13^o34'$E): $V_{r} = 0.3435$ km/s.
The angle $\delta(t) = \omega_{rot} \left( t + \lambda_0 \right)$, where 
$\omega_{rot} = 2\pi/T_d$ with $T_d$ = 1 sidereal day.

\subsection{The time dependence of $ \left| \vec{v}_{lab}(t) \right| $}

In most evaluations of Dark Matter candidates the expected counting rate 
depends on the module of the detector's velocity in the Galaxy, $ v_{lab}(t) $.
The time-independent contribution is $ \left| \vec{v}_{s} \right| = \left| \vec{v}_{LSR} + \vec{v}_{\odot} \right| \approx 232 \pm50$ km/s,
while a Taylor expansion can be performed in the smaller 
time-dependent contributions $ \left| \vec{v}_{rev}(t) \right| \approx 30$ km/s and
$ \left| \vec{v}_{rot}(t) \right| \approx 0.34$ km/s.
Thus, to the first order, that reads:
\begin{equation}
\label{eq:vlab}
 v_{lab}(t) \simeq v_s + \hat{v}_s \cdot \vec{v}_{rev}(t) + \hat{v}_s \cdot \vec{v}_{rot}(t) .
\end{equation}
The higher order terms with no time dependence and with higher harmonics of $\vec{v}_{rev}(t)$:
$ \frac{1}{2} \frac{V_{Earth}^2}{\left| \vec{v}_s \right|} - 
\frac{1}{2} \frac{\left( \hat{v}_s \cdot \vec{v}_{rev}(t) \right)^2}{\left| \vec{v}_s \right|}$,
and the contributions arising from the ellipticity of the Earth orbit
are omitted for simplicity, having frequencies well separated from the diurnal one.

The second term in eq. \ref{eq:vlab} is responsible for the DM annual modulation of the signal and can be written as:
$$
\hat{v}_s \cdot \vec{v}_{rev}(t) = V_{Earth}(\hat{v}_s \cdot \hat{e}^{ecl}_1\sin\lambda(t) - \hat{v}_s \cdot \hat{e}^{ecl}_2\cos\lambda(t)).
$$

\begin{figure}[!h]
\begin{center}
\includegraphics[width=0.49\textwidth]{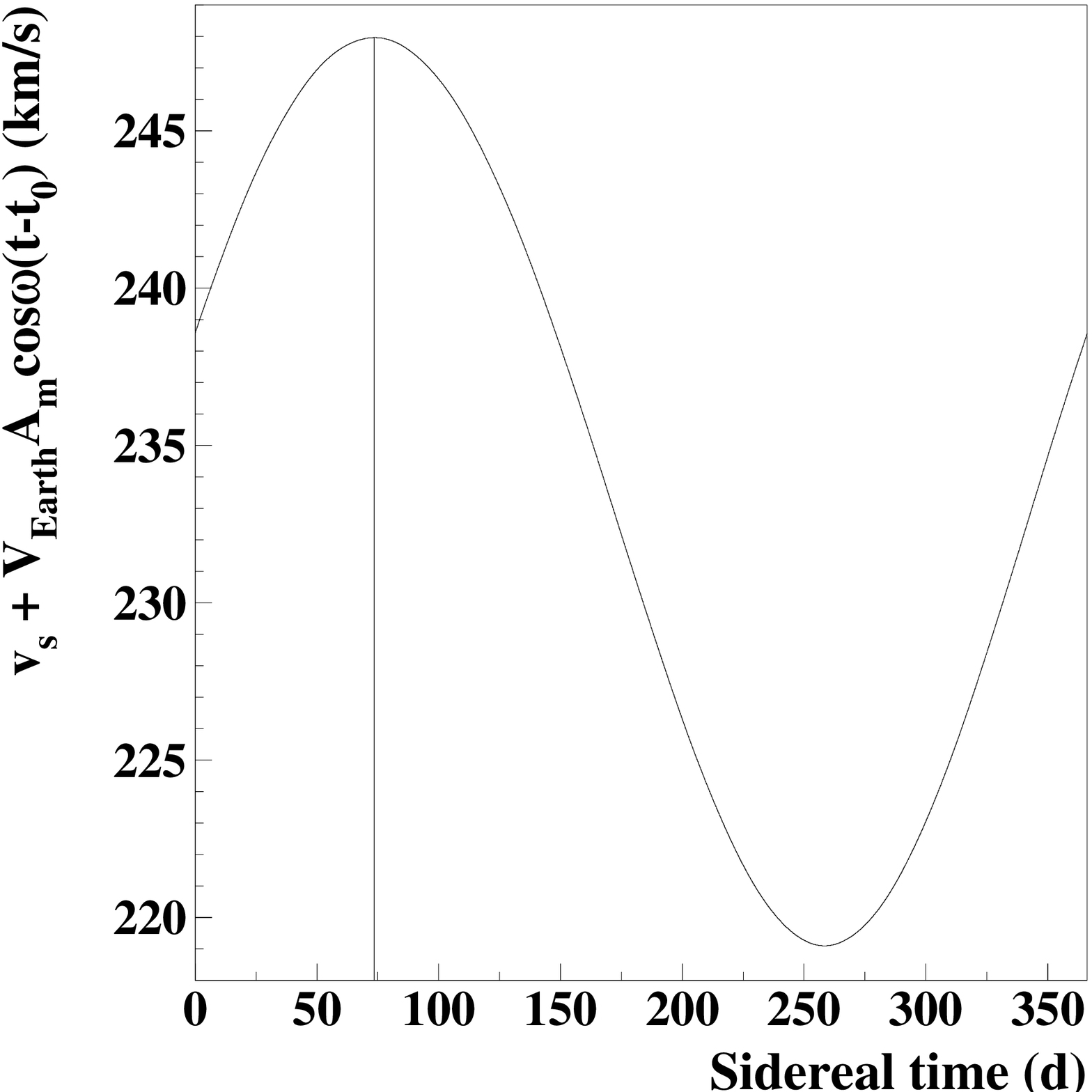}
\includegraphics[width=0.49\textwidth]{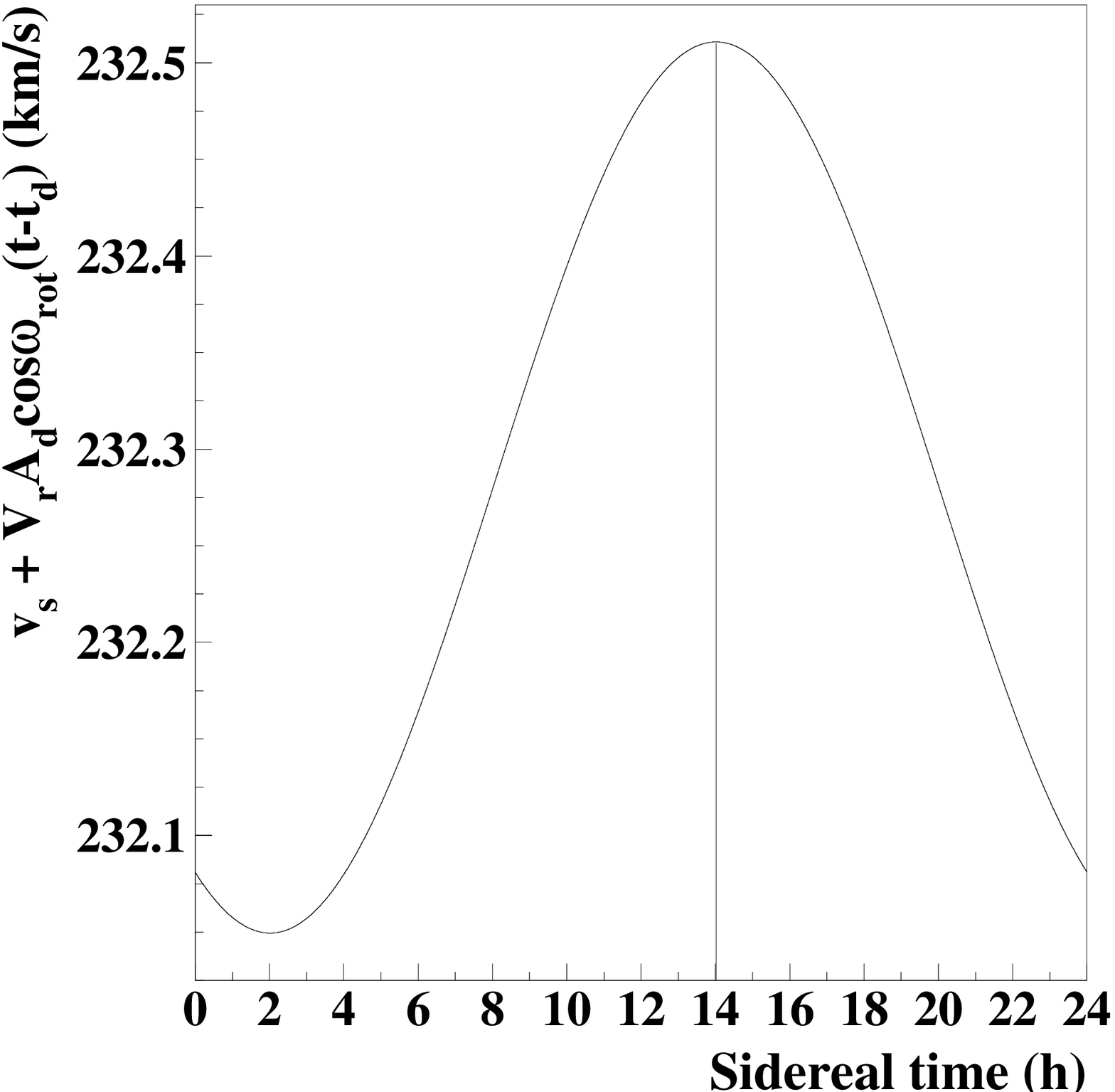}
\end{center}
\vspace{-0.5cm}
\caption{{\it Left:} velocity of the Earth in the galactic frame as a function of the sidereal time,
with starting point March 21 (around spring equinox).
The contribution of diurnal rotation (the third term in eq. \ref{eq:vlab}) has been dropped off.
The maximum of the velocity (vertical line) is about 73 days after the spring equinox.
{\it Right:} sum of the Sun velocity in the galactic frame ($v_s$) and of the rotation velocity of a detector at LNGS 
($\hat{v}_s \cdot \vec{v}_{rot}(t)$) as a function of the sidereal time. 
The maximum of the velocity is about at 14 h (vertical line).
These velocities have been calculated assuming $v_0=220$ km/s by using the routines of Ref. \cite{starlink}.}
\label{phase}
\end{figure}

Defining $\hat{v}_s \cdot \hat{e}^{ecl}_1 = A_m sin \beta_m $ and
$-\hat{v}_s \cdot \hat{e}^{ecl}_2 = A_m cos \beta_m $
(equal to 0.465 and to 0.149, respectively, for the case $v_0=220$ km/s), one can write:
$$
\hat{v}_s \cdot \vec{v}_{rev}(t) = V_{Earth} A_m \cos(\lambda(t) - \beta_m) = V_{Earth} A_m \cos(\omega(t - t_0)),
$$
with $A_m \simeq 0.489$, this confirms that the ecliptic is tilted with respect to the Galactic
plane of $\simeq 60^o$, and $\beta_m \simeq 1.260$ rad (both values are calculated for $v_0=220$ km/s).
The phase of the DM annual modulation is determined at the time when the argument of cosine is null:
$$
 t_0 = t_{equinox} + \beta_m/\omega = t_{equinox} + 73.25 \textrm{ solar days},
$$
for $v_0=220$ km/s; it corresponds to $\approx$ June 2nd (see Fig. \ref{phase}--{\it left}). 
The term in the previous equation ranges from 71.76 solar days (for $v_0=170$ km/s)
to 74.20 solar days (for $v_0=270$ km/s).

The same procedure can be followed to determine the phase of the diurnal modulation due to the
Earth rotation around its axis, described by the third term in eq. \ref{eq:vlab}:
$$
\hat{v}_s \cdot \vec{v}_{rot}(t) = -V_{r}(\hat{v}_s \cdot \hat{e}^{ecs}_1\sin\delta(t) - \hat{v}_s \cdot \hat{e}^{ecs}_2\cos\delta(t)).
$$
Defining $\hat{v}_s \cdot \hat{e}^{ecs}_1 = -A_d sin \beta_d$ and
$\hat{v}_s \cdot \hat{e}^{ecs}_2 = A_d cos \beta_d$
(equal to 0.465 and to -0.484, respectively, for the case $v_0=220$ km/s), one can write:
$$
\hat{v}_s \cdot \vec{v}_{rot}(t) = V_{r} A_d \cos(\delta(t) - \beta_d) = V_{r} A_d \cos\left[\omega_{rot}\left(t - t_{d}\right)\right],
$$
with $A_d \simeq 0.671$ and $\beta_d \simeq 3.907$ rad (both values are calculated for $v_0=220$ km/s).
The phase of the DM diurnal modulation is determined at the time when the argument of cosine is null:
$$
 t_{d} = \beta_d/\omega_{rot} - \lambda_0.
$$
It corresponds to $t_{d} \simeq 14.02$ h sidereal time for the case of a detector at the Gran Sasso longitude\footnote{Note that
in terms of local sidereal time the phase of the DM diurnal modulation is given by $t_d + \lambda_0 = \beta_d/\omega_{rot} \simeq 14.92$ h 
and it is the same for each laboratory independently of its longitude.}
and $v_0=220$ km/s (see Fig. \ref{phase}--{\it right}); actually this value ranges from 13.94 h ($v_0=170$ km/s) 
to 14.07 h ($v_0=270$ km/s).

\vspace{0.3cm}

Finally, the detector's velocity in the Galaxy can be written as:

\begin{equation}
\label{eq:vlab_fin}
 v_{lab}(t) \simeq v_s + V_{Earth} A_m \cos\left[\omega(t - t_0)\right] + V_{r} A_d \cos\left[\omega_{rot}\left(t - t_{d}\right)\right].
\end{equation}

\subsection{The time dependence of the counting rate}

In most evaluations of Dark Matter candidates the expected counting rate 
depends on the module of the detector's velocity in the Galaxy, $ v_{lab}(t) $.
Applying a Taylor expansion, as done in the previous Section, 
the expected signal counting rate in a given $k-th$ energy bin can be written as:
\begin{equation}
S_k\left[ v_{lab}(t) \right] \simeq S_k\left[ v_s \right] + \left[ \frac{\partial  S_k}{\partial v_{lab}} \right] _{v_s}
\left[ V_{Earth} A_m \cos \omega(t - t_0) \right.
+ \left. V_{r} A_d \cos \omega_{rot}\left(t - t_{d}\right) \right].
\label{eq:sm2}
\end{equation}

\noindent The higher order terms with no time dependence and with higher harmonics of $\omega$ are omitted
for simplicity.
In eq. \ref{eq:sm2} the first term is the constant part of the signal ($S_0$), the second term is the 
annual modulation term with amplitude $S_m=\left[ \frac{\partial  S_k}{\partial v_{lab}} \right] _{v_s} V_{Earth} A_m $,
and the third term provides a diurnal modulation with amplitude 
$S_d=\left[ \frac{\partial  S_k}{\partial v_{lab}} \right] _{v_s} V_{r} A_d $.

The interest in this signature is that the ratio $R_{dy}$ of this diurnal modulation amplitude over the annual 
modulation amplitude is a model independent constant; considering the LNGS latitude one has:

\begin{equation}
R_{dy} = \frac{S_d}{S_m} = \frac{ V_{r} A_d }{ V_{Earth} A_m } \simeq 0.016
\label{eq:sm3}
\end{equation}

Taking into account $R_{dy}$ and the annual modulation effect evidenced by DAMA/LIBRA--phase1 for {\it single-hit} events 
in the low energy region, it is possible to derive the diurnal modulation amplitude expected for the same data.
In particular, when considering the (2--6) keV energy interval, the observed annual modulation amplitude in DAMA/LIBRA--phase1 is: 
(0.0097 $\pm$ 0.0013) cpd/kg/keV \cite{modlibra3} and the expected value of the diurnal modulation amplitude is 
$\simeq 1.5\times$ 10$^{-4}$ cpd/kg/keV.

\section{The experimental set-up}
\label{s:setup}

The results presented in the following have been obtained by analysing the data collected in 7 annual cycles by
DAMA/LIBRA--phase1 (1.04 ton$\times$yr exposure) \cite{modlibra,modlibra2,modlibra3} at LNGS.
The description, radiopurity and main features of the DAMA/LIBRA--phase1 setup are discussed in details in the dedicated 
Ref. \cite{perflibra}.
The sensitive part is made of 25 highly radiopure NaI(Tl) crystal scintillators organized in a (5 $\times$ 5) matrix; each NaI(Tl) 
detector has 9.70 kg mass and a size of ($10.2 \times 10.2 \times 25.4$) cm$^{3}$.
The bare crystals are enveloped in Tetratec-teflon foils and encapsulated in radiopure OFHC Cu housing.
In each detector two 10 cm long special quartz light guides act also as optical windows on the two end faces of the crystal 
and are coupled to two low background photomultipliers (PMT) working in coincidence at single photoelectron level.
The detectors are housed in a sealed low-radioactive copper box installed in the center of a low-radioactive 
Cu/Pb/Cd-foils/polyethylene/paraffin shield; moreover, about 1 m concrete (made from the Gran Sasso rock material) 
almost fully surrounds (mostly outside the barrack) this passive shield, acting as a further neutron moderator.
The copper box is maintained in HP Nitrogen atmosphere in slightly overpressure with respect to the external 
environment; it is part of the threefold-level sealing system which excludes the detectors from the environmental air 
of the underground laboratory.
The light response of the detectors in DAMA/LIBRA--phase1 typically ranges from 5.5 to 7.5 
photoelectrons/keV, depending on the detector.
The hardware threshold of each PMT is at single photoelectron, while a software energy threshold of 2 keV electron 
equivalent (hereafter keV) is used \cite{perflibra}.
Energy calibration with X-rays/$\gamma$ sources are regularly carried out in the same running condition down to few 
keV \cite{perflibra}. Moreover, the whole DAMA/LIBRA installation is under air conditioning to assure a suitable and stable 
working temperature for the electronics; in addition, the huge heat capacity of 
the multi-ton passive shield ($\simeq 10^6$ cal/$^o$C) further 
assures a relevant stability of the detectors' operating temperature. 
The DAQ system records both {\it single-hit} events (where just one of the detectors fires) and {\it multiple-hit} 
events (where more than one detector fire) up to the MeV region despite the optimization is performed for the lowest 
one. 
A hardware/software system is operative to monitor the running conditions, and self-controlled computer processes 
automatically control several parameters and manage alarms.
For the radiopurity, the electronic chain, the data acquisition system and for all the other details see Ref. 
\cite{perflibra}; for completeness we recall that during DAMA/LIBRA--phase1 new transient
digitizers and DAQ system have been installed at fall 2008 before the start of the sixth annual
cycle.

\section{Model independent experimental results}
\label{s:resid}

In order to point out the presence of a possible diurnal effect, the
low energy {\it single-hit} DAMA/LIBRA--phase1 data have been 
grouped in 1 hour bin using either the sidereal or solar time, respectively.
We define $N_{i,d}^{(jky)}$ as the number of events collected in the $i$th hour of the day $d$
(for the two cases of sidereal and solar time); as regards the other indexes: 
i) $j$ identifies the detector; 
ii) $k$  identifies the energy bin within the considered energy interval; 
iii) $y$ identifies the annual cycle.
Hence, the {\it single-hit} rate in $i$th hour is written as:
$$
r_{i}^{(jky)} = \frac{\sum_d N_{i,d}^{(jky)}}{\sum_d M_j \Delta t_{i,d}^{(y)} \Delta E \epsilon^{(jky)}},
$$
where $M_j$ is the mass of the $j$th detector, $\Delta t_{i,d}^{(y)}$ is the detector running time during the 
$i$th hour of the $d$th day of the $y$th annual cycle, $\Delta E$ is the energy bin and
$\epsilon^{(jky)}$ is the overall efficiency \cite{perflibra}.

Therefore, the residual rate has been calculated according to:
$ \langle r_{i}^{(jky)} - flat^{(jky)} \rangle_{jky}$, 
where the average is made on all the detectors ($j$ index), on all the considered energy bins ($k$ index), and 
on all the DAMA/LIBRA--phase1 annual cycles ($y$ index).
The $flat^{(jky)}$ is the rate averaged over the day: $ flat^{(jky)} = \langle r_{i}^{(jky)} \rangle_i$,
which is of the order of $\approx 1$ cpd/kg/keV \cite{perflibra}.
Here, the possible time-dependent contribution due to the annual modulation signal is not considered since it is
washed out by the almost uniform data collection along the day and along the year; its 
estimated effect is much lower than $\simeq 10^{-4}$ of $flat^{(jky)}$.

\begin{figure}[!p]
\begin{center}
\includegraphics[width=5.2cm]{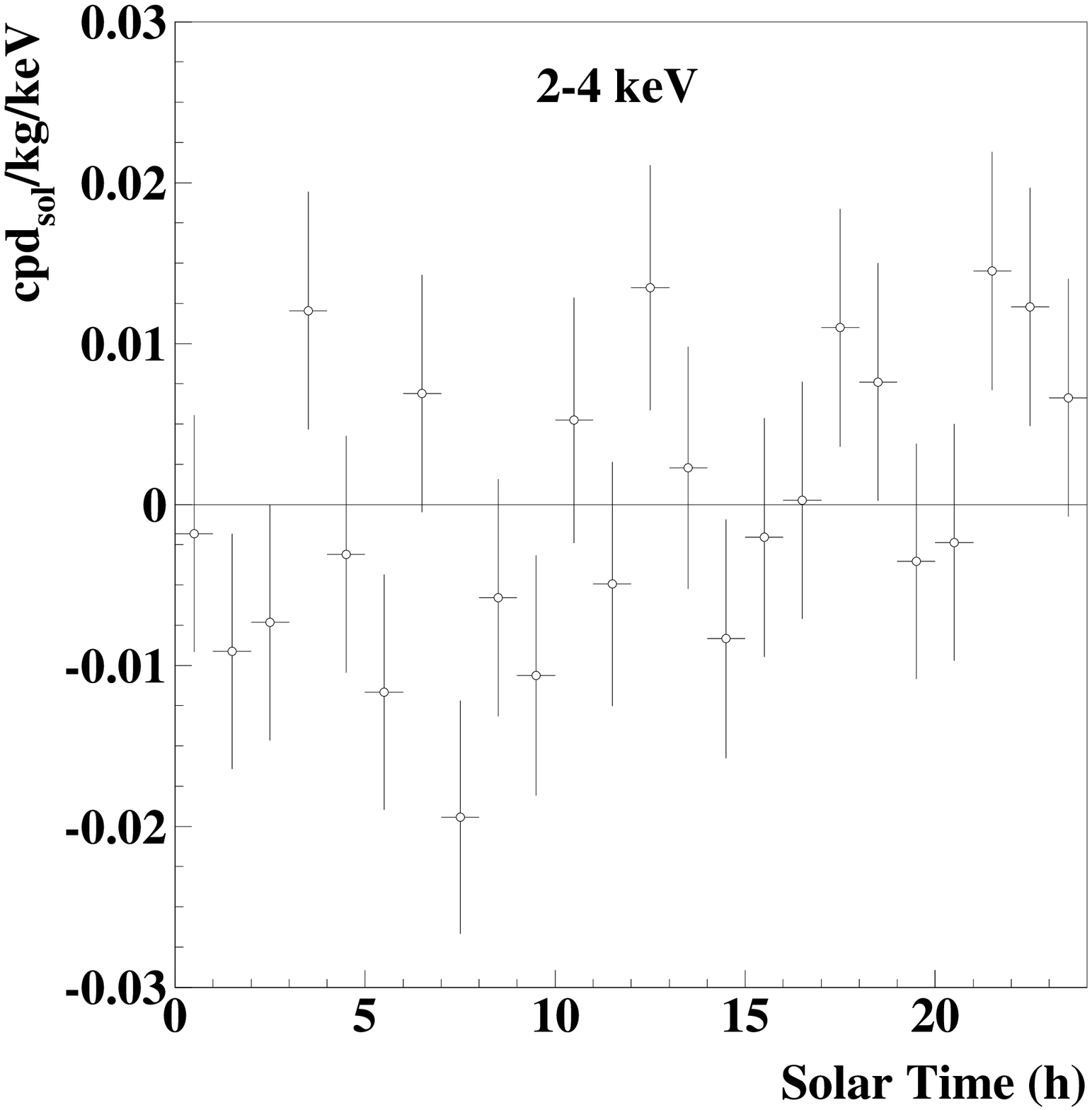}
\includegraphics[width=5.2cm]{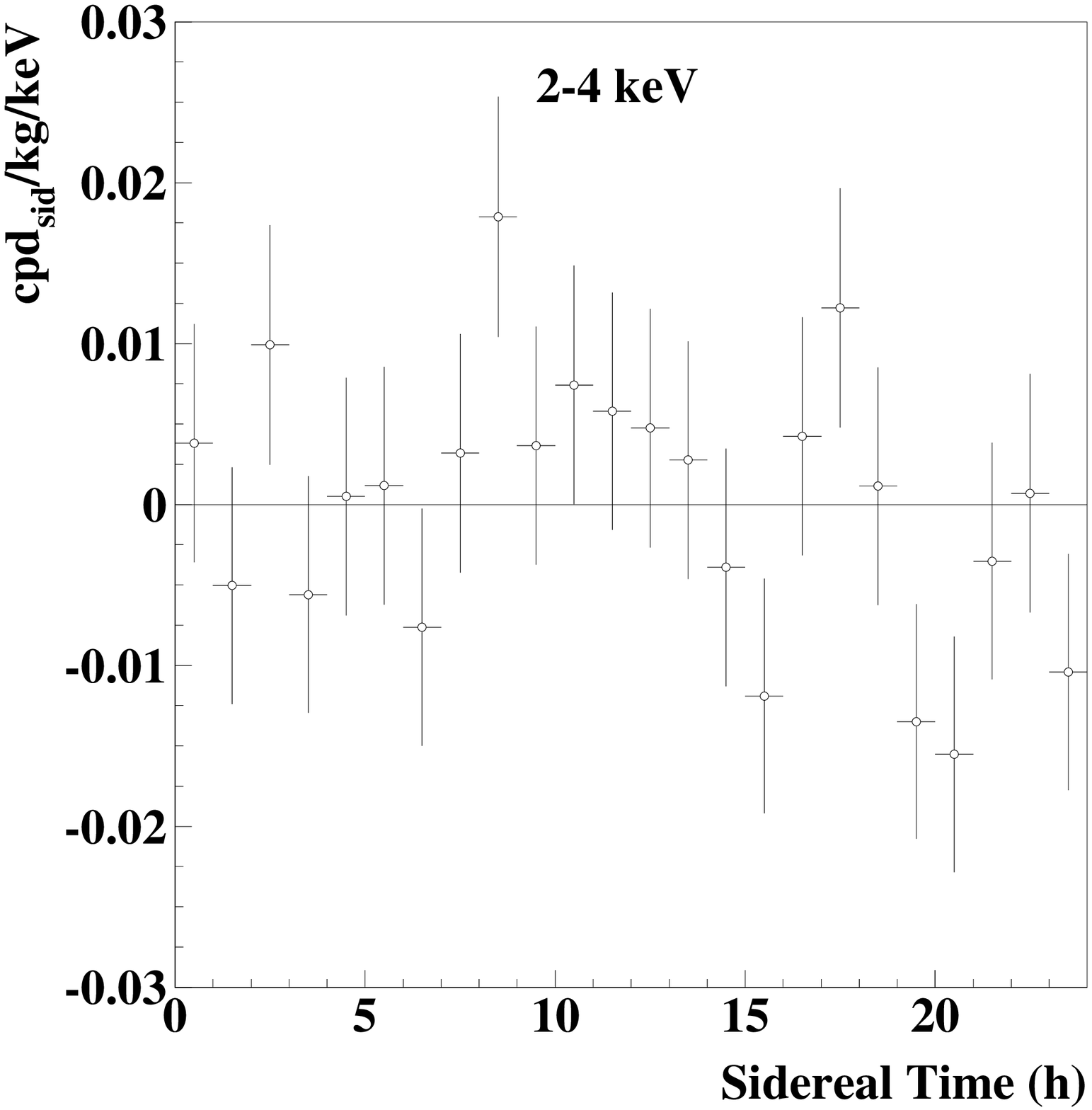}
\includegraphics[width=5.2cm]{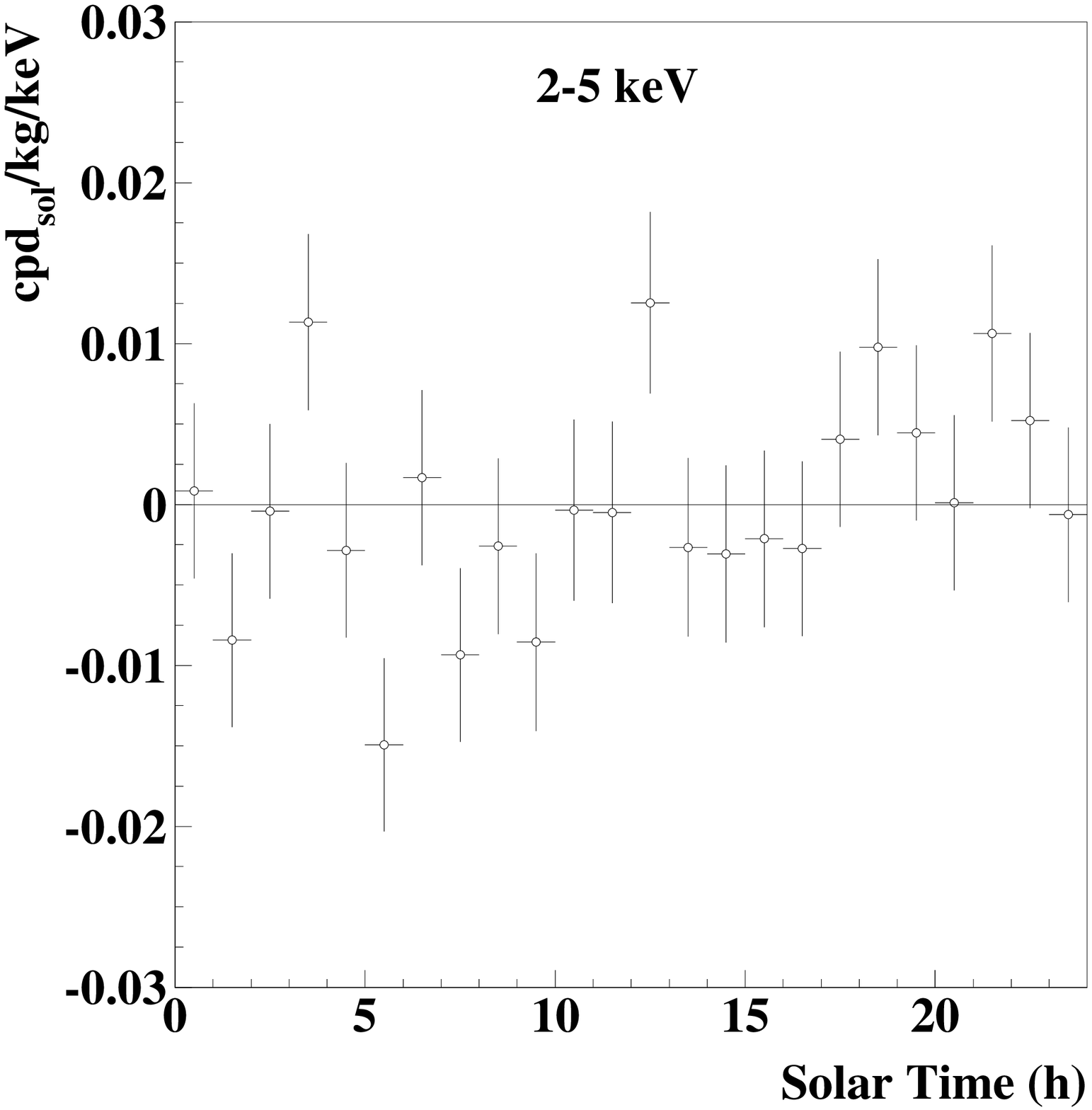}
\includegraphics[width=5.2cm]{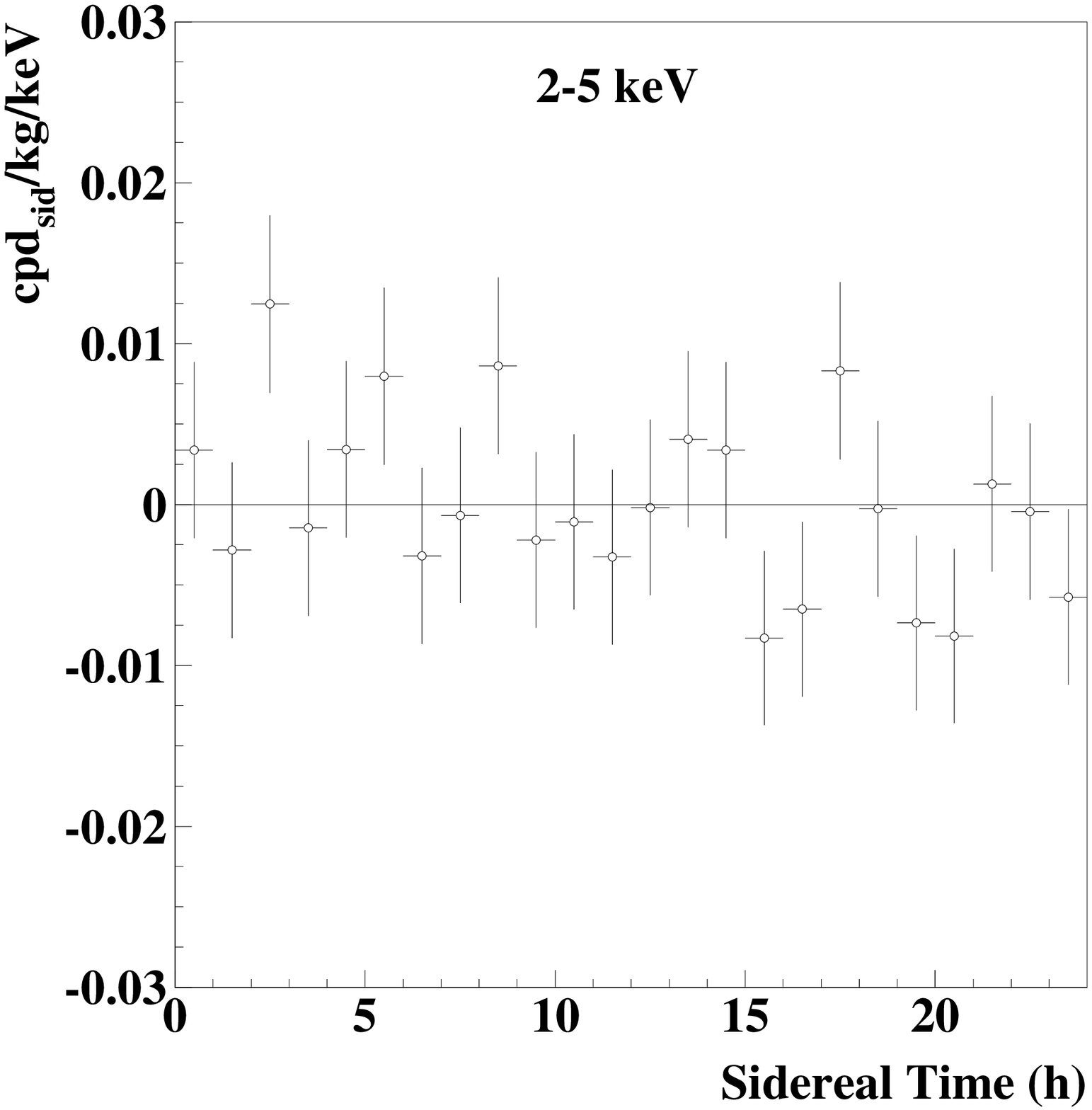}
\includegraphics[width=5.2cm]{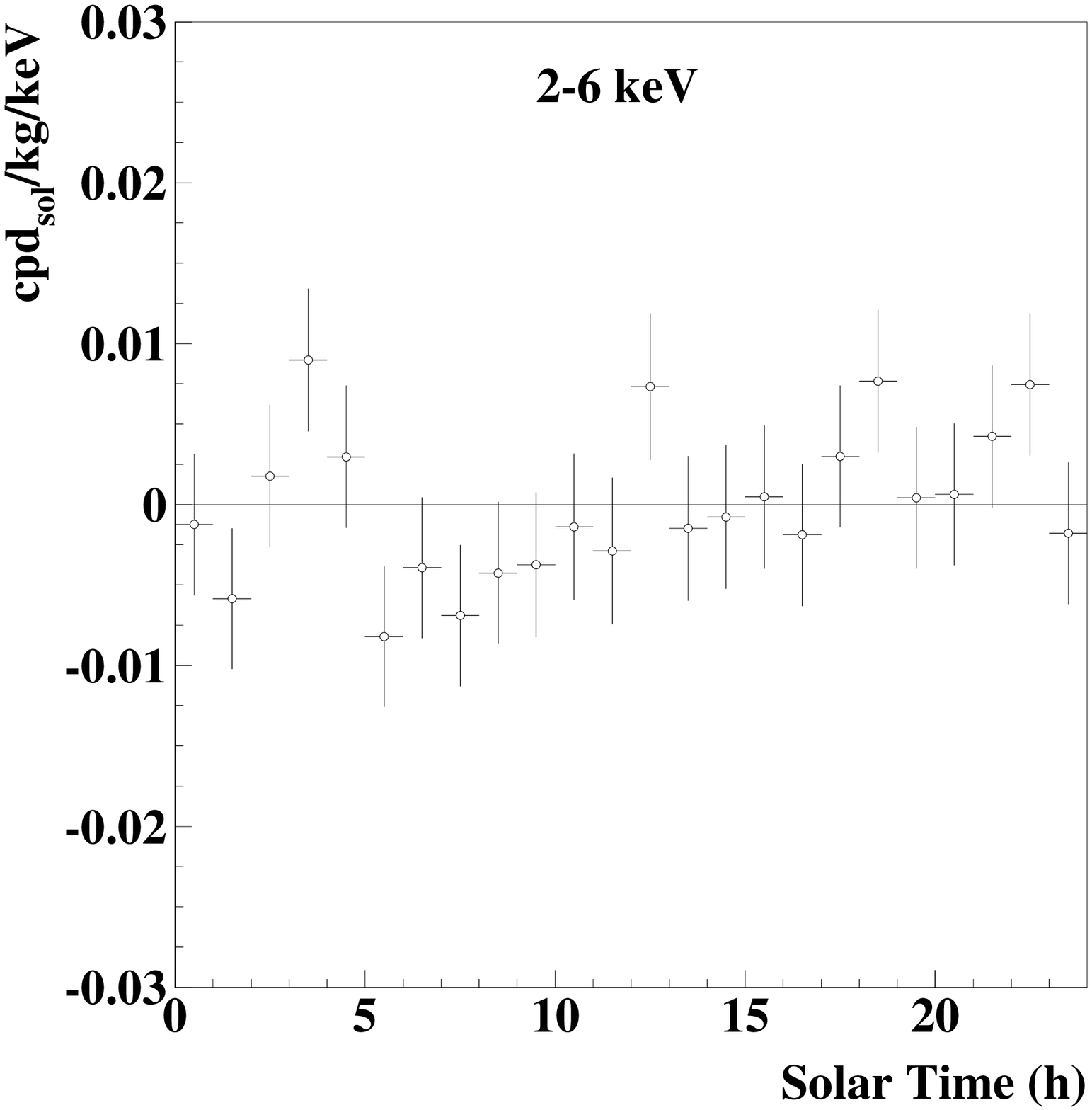}
\includegraphics[width=5.2cm]{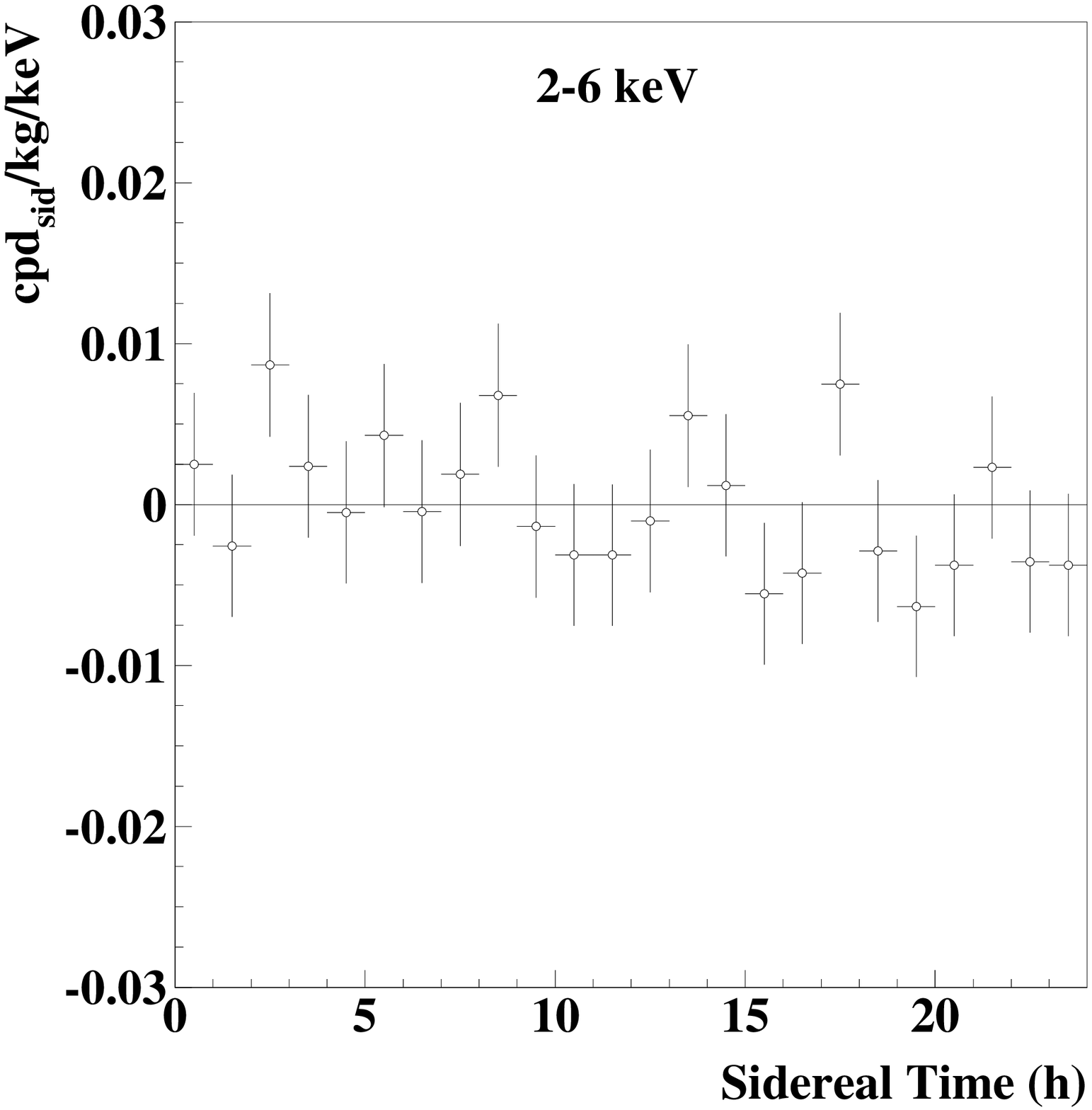}
\end{center}
\caption{
Experimental model-independent diurnal residual rate of the {\it single-hit} scintillation events, measured by 
DAMA/LIBRA--phase1 in the (2--4), (2--5) and (2--6) keV energy intervals as a function 
of the hour of the solar ($left$) and sidereal ($right$) day.
The experimental points present the errors as vertical bars and the associated time bin width 
(1 hour) as horizontal bars. The cumulative exposure is 1.04 ton $\times$ yr. See text.}
\label{resid1}
\end{figure}
\begin{figure}[!ht]
\begin{center}
\includegraphics[width=5.2cm]{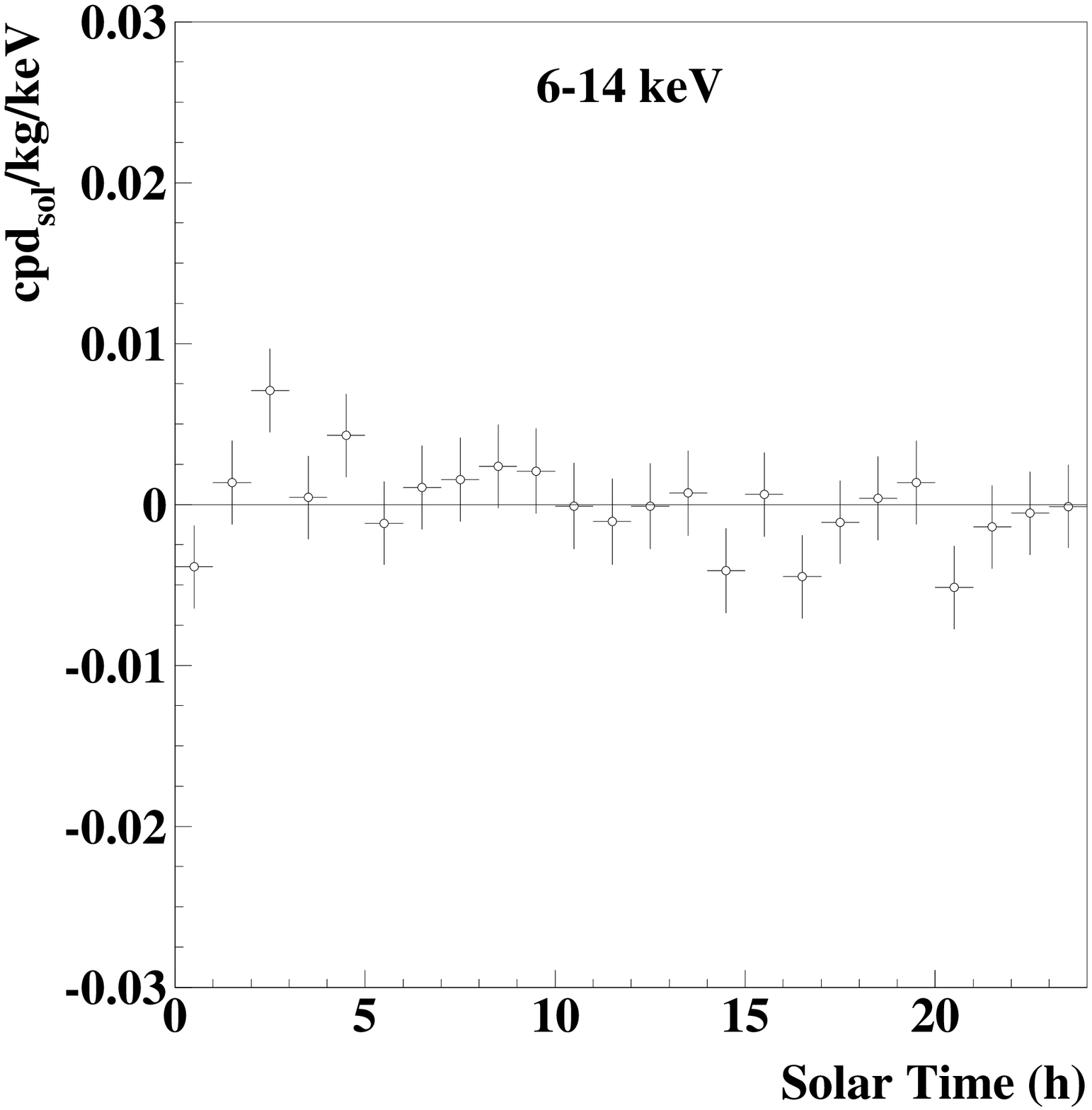}
\includegraphics[width=5.2cm]{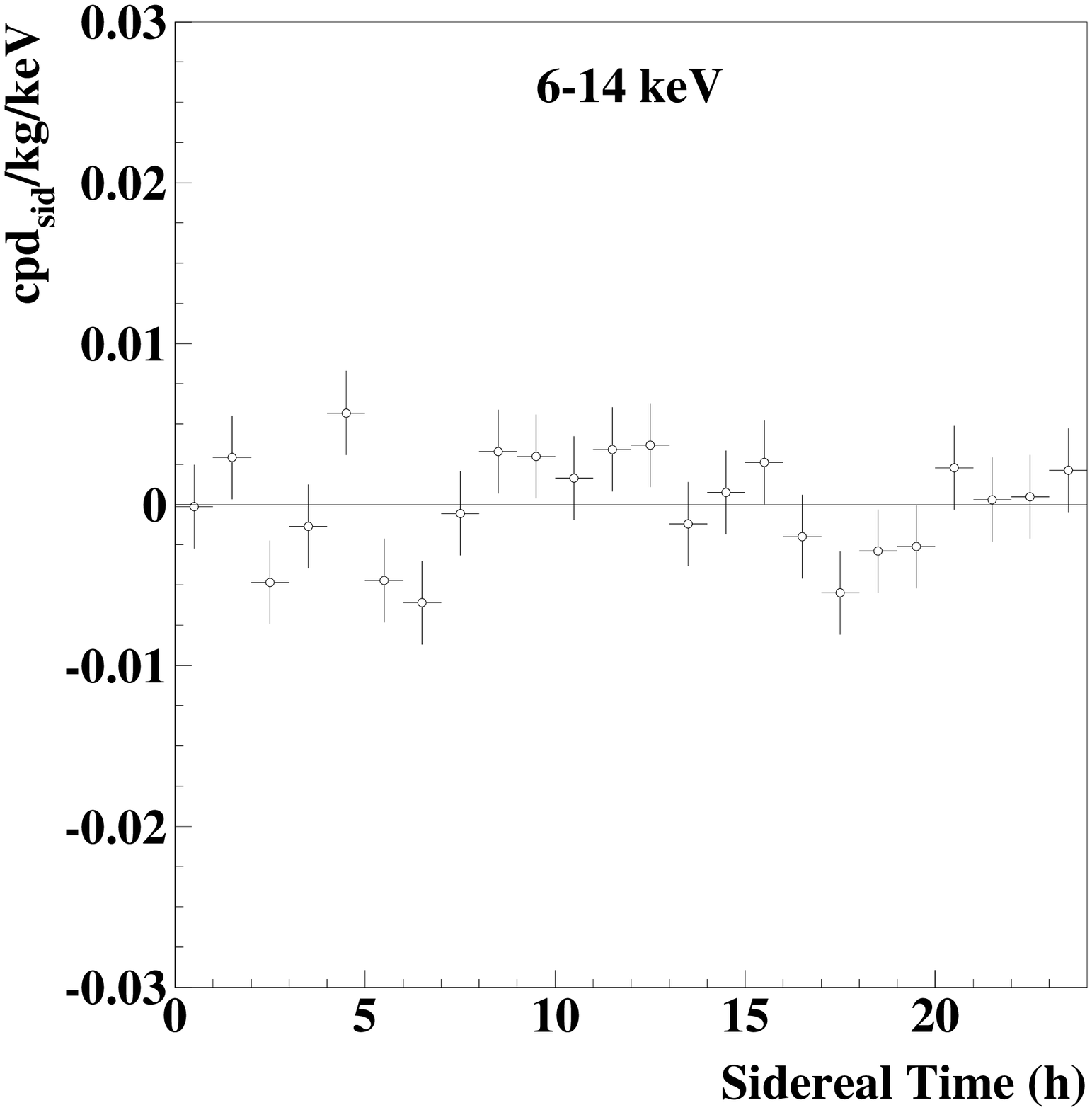}
\end{center}
\caption{
Experimental model-independent diurnal residual rate of the {\it single-hit} scintillation events, measured by 
DAMA/LIBRA--phase1 in the (6--14) keV energy interval as a function of the hour of the solar ($left$) and 
sidereal ($right$) day.
The experimental points present the errors as vertical bars and the associated time bin width 
(1 hour) as horizontal bars. The cumulative exposure is 1.04 ton $\times$ yr. See text.}
\label{resid2}
\vspace{-0.1cm}
\end{figure}

Fig. \ref{resid1} shows the time and energy behavior of the experimental residual rates of {\it single-hit} events  
both as a function of solar ($left$) and of sidereal ($right$) time, in the (2--4), (2--5) and (2--6) keV   
energy intervals \footnote{We recall that the annual modulation signal has been pointed out only in these energy intervals; 
see Ref. \cite{modlibra,modlibra2,modlibra3} and references therein.} and in Fig. \ref{resid2} those in the 
(6--14) keV interval for the solar ($left$) and sidereal ($right$) case.
The used time bin is 1 (either solar or sidereal, respectively) hour. 

The null hypothesis (absence of residual rate diurnal variation) has been tested by a $\chi^2$ test,
obtaining the results given in Table \ref{tb1}; there the upper tail probabilities (P-values),
calculated by the standard $\chi^2$ distribution, are also reported. Thus, no diurnal variation with
a significance of 95\% C.L. is found.

\begin{table}[!h]
\caption{Test of absence of diurnal effect in the DAMA/LIBRA--phase1 data.
The P-values, calculated by the standard $\chi^2$ distribution, are also shown. 
As it can be seen, the $\chi^2$ test applied to the data supports the hypothesis that 
the residual rates are simply fluctuating around zero.}
\begin{center}
\begin{tabular}{|r|l|l|}
\hline
   Energy  &                  Solar Time                      &                  Sidereal Time                   \\ \hline \hline
 2--4  keV & $\chi^2$/d.o.f. = 35.2/24 $\rightarrow$ P =  7\% & $\chi^2$/d.o.f. = 28.7/24 $\rightarrow$ P = 23\% \\ \hline
 2--5  keV & $\chi^2$/d.o.f. = 35.5/24 $\rightarrow$ P =  6\% & $\chi^2$/d.o.f. = 24.0/24 $\rightarrow$ P = 46\% \\ \hline
 2--6  keV & $\chi^2$/d.o.f. = 25.8/24 $\rightarrow$ P = 36\% & $\chi^2$/d.o.f. = 21.2/24 $\rightarrow$ P = 63\% \\ \hline
 6--14 keV & $\chi^2$/d.o.f. = 25.5/24 $\rightarrow$ P = 38\% & $\chi^2$/d.o.f. = 35.9/24 $\rightarrow$ P =  6\% \\ \hline
\end{tabular}
\end{center}
\vspace{-0.6cm}
\label{tb1}
\end{table}

In addition to the $\chi^2$ test, another independent statistical test has been applied: the run test (see e.g. Ref. \cite{eadie});
it verifies the hypothesis that the positive and negative data points are randomly distributed.
The lower tail probabilities are equal to: 
43\%, 18\%,  7\% and 26\% in the (2--4), (2--5), (2--6) and (6-14) keV energy region, respectively, for the solar case and 
54\%, 84\%, 78\% and 16\% in the (2--4), (2--5), (2--6) and (6-14) keV energy region, respectively, for the sidereal case.
Thus, in conclusion the presence of any significant diurnal variation and of time structures can be excluded at the reached 
level of sensitivity (see e.g. the error bars in Fig. \ref{resid1}).

\subsection{Comparison with expectation for DM diurnal effect}

When considering the DM diurnal effect due to the Earth rotation around its axis described 
in Sect. \ref{s:diurnal}, only an upper limit can be derived. 
In particular, the residual rates of the $single$-$hit$ events in the (2--4), (2--5), (2--6) and (6--14) keV 
energy intervals as a function of the sidereal time 
(see Figs. \ref{resid1} $right$ and \ref{resid2} $right$) 
have been fitted with a cosine function with amplitude $A_d^{exp}$ as free parameter, period 
fixed at 24 h and phase at 14 h. The results are reported in Table \ref{tb2}: all the 
diurnal modulation amplitudes are compatible with zero. 

\begin{table}[ht]
\caption{Diurnal modulation amplitudes, $A_d^{exp}$, for each considered energy interval 
obtained by fitting the {\it single-hit}
residual rate of the entire DAMA/LIBRA--phase1 as function of the sidereal time, 
(see Figs. \ref{resid1} $right$ and \ref{resid2} $right$) with the formula
$A_d^{exp} \cos\left[\omega_{rot}\left(t - t_{d}\right)\right]$.
The amplitude $A_d^{exp}$ is a free parameter, while the period 
is fixed at 24 h and the phase at 14 h, as expected for the DM diurnal effect.
The corresponding $\chi^2$ values of each fit and the P-values are also reported.}
\begin{center}
\begin{tabular}{|r|c|c|c|}
\hline
   Energy  & $A_d^{exp}$ (cpd/kg/keV)        &   $\chi^2$/d.o.f.  &   P      \\ \hline \hline
 2--4  keV & $ (2.0 \pm 2.1) \times 10^{-3}$ &   27.8/23          &  22\%    \\ \hline
 2--5  keV & $-(1.4 \pm 1.6) \times 10^{-3}$ &   23.2/23          &  45\%    \\ \hline
 2--6  keV & $-(1.0 \pm 1.3) \times 10^{-3}$ &   20.6/23          &  61\%    \\ \hline
 6--14 keV & $ (5.0 \pm 7.5) \times 10^{-4}$ &   35.4/23          &   5\%    \\ \hline
\end{tabular}
\end{center}
\label{tb2}
\end{table}

\begin{figure}[!ht]
\begin{center}
\includegraphics[width=0.92\textwidth]{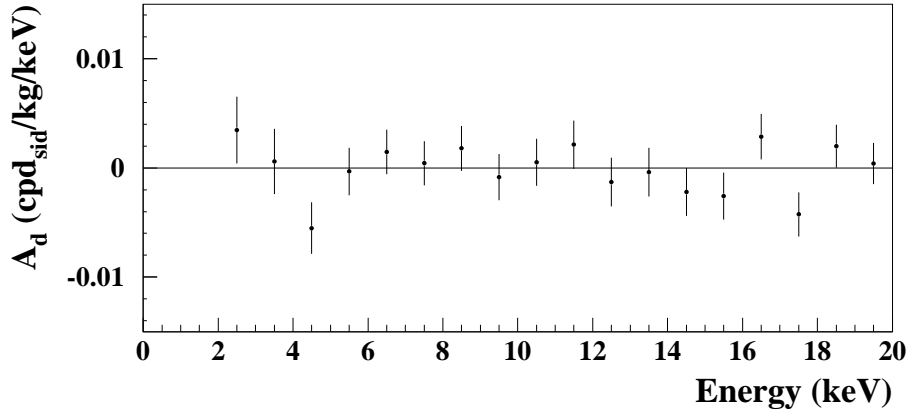}
\end{center}
\caption{Diurnal modulation amplitudes, $A_d$, as function of energy (the energy bin is 1 keV)
obtained by fitting the {\it single-hit}
residual rate of the entire DAMA/LIBRA--phase1 as function of the sidereal time, 
with the formula $A_d \cos\left[\omega_{rot}\left(t - t_{d}\right)\right]$.
The amplitude $A_d$ is a free parameter, while the period 
is fixed at 24 h and the phase at 14 h, as expected for the DM diurnal effect.
The $A_d$ values are compatible with zero, having 
random fluctuations around zero with $\chi^2$ equal to 19.5 for 18 degrees of freedom.
The cumulative exposure is 1.04 ton $\times$ yr. See text.}
\label{fg:amp_sid}
\end{figure}

\vspace{0.3cm}
Fig. \ref{fg:amp_sid} shows the diurnal modulation amplitudes, $A_d$, as function of energy (the energy bin is 1 keV)
obtained by fitting the {\it single-hit}
residual rate of the entire DAMA/LIBRA--phase1 as function of the sidereal time, 
with the formula $A_d \cos\left[\omega_{rot}\left(t - t_{d}\right)\right]$.
The period is fixed at 24 h and the phase at 14 h, as expected for the DM diurnal effect (see above).
The $A_d$ values are compatible with zero, having 
random fluctuations around zero with $\chi^2$ equal to 19.5 for 18 degrees of freedom.

\vspace{0.3cm}
In order to compare the experimental data with the DM diurnal effect due to the Earth rotation around its axis described 
in Sect. \ref{s:diurnal}, the (2--6) keV energy interval is taken into account for simplicity.
From Table \ref{tb2} one can obtain 
$A_d^{exp} = -(1.0 \pm 1.3) \times 10^{-3}$ cpd/kg/keV ($\chi^2/d.o.f. = 20.6/23$).
Following the Feldman-Cousins \cite{FeCo} procedure an upper limit can be obtained for the measured diurnal modulation 
amplitude: $A_d^{exp} < 1.2 \times 10^{-3}$ cpd/kg/keV (90\% C.L.); thus, the present experimental sensitivity is larger 
than the expected diurnal modulation amplitude ($\simeq 1.5\times$ 10$^{-4}$ cpd/kg/keV) derived above from the 
DAMA/LIBRA--phase1 observed effect.

\vspace{0.3cm}
In conclusion, it will be possible to investigate this diurnal effect with adequate sensitivity 
only when a much larger exposure will be available,
provided a suitable control of the running parameters at the needed level.
On the other hand, better sensitivities can also be achieved by lowering the software energy threshold;
in fact an almost exponential rising of the signal rate is expected at lower energy for some DM candidates.
This is one of the goals of the presently running DAMA/LIBRA--phase2.

\vspace{0.5cm}
\subsection{Comparison with any hypothetical diurnal effects with cosine behaviour}

\begin{figure}[!ht]
\begin{center}
\includegraphics[width=0.7\textwidth]{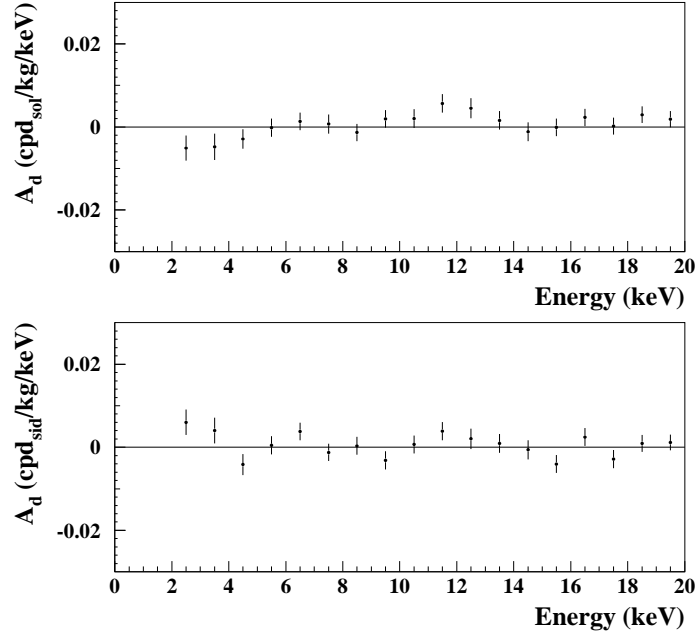}
\end{center}
\vspace{-0.5cm}
\caption{Diurnal modulation amplitudes, $A_d$, as function of energy (the energy bin is 1 keV)
obtained by fitting the {\it single-hit}
residual rate of the entire DAMA/LIBRA--phase1 
with the formula $A_d \cos\left[\omega_{rot}\left(t - t_{d}^*\right)\right]$.
The free parameters of the fit are the 18 modulation amplitudes (one for each energy bin)
and the phase $t_d^*$. The results are reported for
the solar time ($top$) and sidereal time ($bottom$) cases.
The $A_d$ values are compatible with zero, having 
random fluctuations around zero with $\chi^2$ equal to 24.2 and 25.4 (18 degrees of freedom)
for the solar time and sidereal time, respectively.
The best fit values for the phase are $t_d^* = (6.1 \pm 1.1)$ h and $t_d^* = (10.7 \pm 1.1)$ h 
for the solar time and sidereal time, respectively.
The cumulative exposure is 1.04 ton $\times$ yr. See text.}
\label{fg:amp_tvar}
\end{figure}

In order to leave to the reader the possibility to compare the data with 
possible exotic models,
the experimental residual rates of the {\it single-hit} events 
as function of both solar and sidereal time have been compared 
with a cosine function with a free phase. For this purpose 
the residual rate for each energy bin of 1 keV has been fitted 
with the formula $A_d \cos\left[\omega_{rot}\left(t - t_{d}^*\right)\right]$.
The free parameters of the fit are the 18 modulation amplitudes (one for each energy bin)
and the phase $t_d^*$. The period is fixed at 24 h. 
The results are reported in Fig. \ref{fg:amp_tvar} for 
both solar and sidereal time cases. 

The $A_d$ values are compatible with zero, having 
random fluctuations around zero with $\chi^2$ equal to 24.2 and 25.4 (18 degrees of freedom)
for the solar time and sidereal time, respectively.
The best fit values for the phase are $t_d^* = (6.1 \pm 1.1)$ h and $t_d^* = (10.7 \pm 1.1)$ h 
for the solar time and sidereal time, respectively.

\subsection{Comparison with possible diurnal effects induced by cosmic rays}

Solar and sidereal diurnal modulation of the underground muon rate at LNGS have been searched for by the MACRO experiment 
computing hourly deviations of the muon rate from 6 month averages \cite{amb03}.
Statistically significant diurnal modulations with the solar and the sidereal periods have been pointed out: their amplitudes
are $<0.1$\%, at the limit of the detector statistics.
The solar diurnal modulation is due to the diurnal atmospheric temperature variations at 20 km, i.e. the altitude of primary cosmic 
ray interactions with the atmosphere; the sidereal diurnal modulation is due to the Compton-Getting modulation due to solar system 
motion relative to the local standard of rest.

Thus, we have estimated consistency of the result, obtained with the presently reached
sensitivity, with these known effects.
The measured {\it single-hit} event counting rate of DAMA/LIBRA--phase1 in the low energy region is of the order of $\approx$ 
1 cpd/kg/keV \cite{perflibra} and,
as shown in Ref. \cite{mu}, the contribution due to muons surviving the Gran Sasso mountain or related particles is very small 
($\ll 1$ cpd/kg/keV).
Therefore, the effect due to diurnal variation of muon flux is expected to be $\ll 10^{-3}$ cpd/kg/keV, 
that is well below the present experimental sensitivity (see e.g. the error bars in Fig. \ref{resid1}).

\section{Conclusions}

The low energy (2--6) keV {\it single-hit} data collected in the whole DAMA/LIBRA--phase1 (7 annual cycles; exposure: 1.04 ton $\times$ yr) 
\cite{modlibra,modlibra2,modlibra3} have been analyzed in terms of diurnal effects.
At the present level of sensitivity the presence of any significant diurnal variation and of diurnal time structures in the data can 
be excluded for both the cases of solar and sidereal time. 
In particular, the diurnal modulation amplitude expected -- because 
of the Earth diurnal motion -- on the basis of the DAMA DM annual modulation results is below the present sensitivity;
it will be possible to investigate this diurnal effect with adequate sensitivity only when a much larger exposure will be available,
provided a suitable control of the running parameters at the needed level.
At present DAMA/LIBRA is continuously running in its new configuration (named DAMA/LIBRA--phase2) with a lower software 
energy threshold \cite{pmts} which also can offer an alternative possibility to increase sensitivity to such an effect.

\section{Acknowledgments} 

It is a pleasure to thank Mr. A. Bussolotti and A. Mattei for their qualified technical work.

\end{document}